\documentclass[9pt,twocolumn,twoside]{osajnl}

\journal{ol} 

\setboolean{shortarticle}{true}

\title{946-nm Nd:YAG digital-locked laser at $1.1\times10^{-16}$ in 1~s and transfer-locked to a cryogenic silicon cavity}

\author[1,*]{Alexandre Didier}
\author[2]{Stepan Ignatovich}
\author[1]{Erik Benkler}
\author[1,2]{Maxim Okhapkin}
\author[1]{Tanja E. Mehlstäubler}

\affil[1]{Physikalisch-Technische Bundesanstalt, Bundesallee 100, 38116 Braunschweig, Germany}
\affil[2]{Institute of Laser Physics SB RAS, Novosibirsk, Russia}

\affil[*]{Corresponding author: alexandre.didier@ptb.de}

\begin{abstract}
We present a Nd:YAG ultra-stable laser system operating at 946~nm and demonstrate a fractional frequency instability of $1.1\times10^{-16}$ at 1~s by pre-stabilizing it to a 30~cm-long ULE cavity at room temperature. All key analog components have been replaced by FPGA-based digital electronics. To reach an instability below the $10^{-16}$ level, we transfer the stability of a 1542~nm laser stabilized to a cryogenic silicon cavity exhibiting a fractional frequency instability of $4\times10^{-17}$ at 1~s to the laser at 946~nm.
\end{abstract}

\setboolean{displaycopyright}{true}

\begin{document}

\maketitle


Optical oscillators with high spectral purity have become key components of high-precision experiments, such as gravitational wave detection~\cite{Adhikari_2014} and optical atomic clocks~\cite{Ludlow_2015}.
For the latter, short-term fractional frequency instabilities in the low $10^{-16}$ range or better are required for minimizing the Dick effect in optical lattice clocks~\cite{Hinkley_2013}, achieving long coherent interrogation times in ion clocks~\cite{Ludlow_2015} or to benefit from the low quantum projection noise expected in multi-ions clocks~\cite{Herschbach_2012,Barrett_2015,Keller2019-PRA}. 
Among all available technologies, ultra-stable lasers obtained by referencing a continuous wave (CW) laser to a high finesse Fabry-Perot cavity have been broadly preferred for their reliability and robustness. Many experiments could demonstrate short-term fractional frequency instabilities below $10^{-15}$, but only a few have reached $1\times10^{-16}$ or below to this date. The exceptionally low instabilities were either obtained at room temperature with very long cavities~\cite{Hafner_2015}, which rely on a very carefully designed and optimized vibration-insensitive cavity support, or at cryogenic temperatures with technically more complex systems involving bulky cryocoolers~\cite{Matei2017}.
Furthermore, the traditional electronics involved in the generation of the error signal and the stabilization itself are mostly analog, which have to be carefully designed and are not easy to tune in experiments, e.g. precise filter frequencies, phase adjustment of the Pound-Drever-Hall (PDH) signal, offset drift of the frequency mixer.
In this letter, we present a robust ultra-stable laser based on a solid-state Nd:YAG laser with high control bandwidth operating at 946~nm and stabilized to a 30~cm high finesse Fabry-Perot cavity placed on a simple adjustable vibration-insensitive support~\cite{keller_simple_2014}. 
A similar approach has been followed recently~\cite{LS-Ma_2018} with a 578~nm laser stabilized to two 30~cm-long ULE cavities, where a fractional frequency instability of $2\times10^{-16}$ at 1~s was obtained but is still limited by the vibration sensitivities of the cavities.
In this work, we demonstrate the first frequency stabilization via digital electronics at the fractional instability level of $1.1\times10^{-16}$ at 1~s and discuss the relevant instability contributions.
This laser will be used for high-precision spectroscopy of the ${}^1$S${}_{0}\leftrightarrow {}^3$P${}_{0}$ clock transition of $^{115}$In$^{+}$ in a multi-ion clock~\cite{Herschbach_2012,Keller2019-PRA}, for which it is frequency quadrupled to 236.5~nm.
For reaching even better instabilities, we use a transfer-lock scheme to stabilize the 946~nm laser to a 1542~nm laser locked to a cryogenic silicon cavity with $4\times10^{-17}$ fractional frequency instability from 0.8~s to $\sim$30~s~\cite{Matei2017}.

\label{sec:1-Laser}

\begin{figure}[ht]
	\centering
		\includegraphics[width=0.9\linewidth]{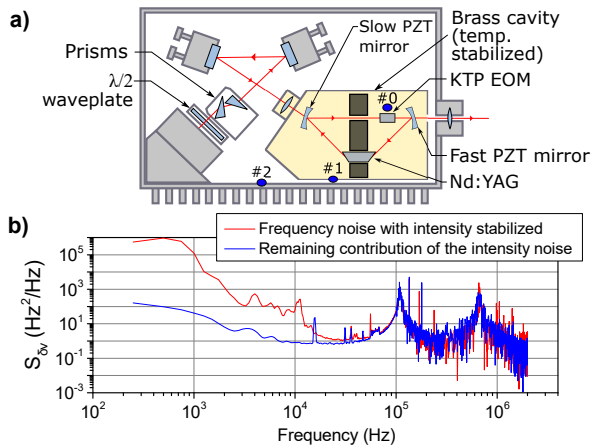}
	\caption{
		\textbf{a} - Design of the Nd:YAG laser at 946~nm. The blue dots represent the temperature sensors used for the active correction of the temperature gradients inside the laser cavity. 
		\textbf{b} - Frequency noise of the free-running laser. The red curve is obtained when the laser intensity is stabilized. The blue curve is the contribution of the remaining intensity noise. The peak at 110~kHz is due to a resonant relaxation oscillation of the Nd:YAG crystal~\cite{Harb:97}. The peak at 700~kHz is a servo bump from the intensity stabilization. 
	}
	\label{fig:Fig1-Laser-scheme-noise}
\end{figure}

The laser used in this work is a home-built quasi-three level solid state Nd:YAG laser operating at 946~nm and is based on the design in~\cite{2001OptCo.194..207O}.
The laser is depicted in figure~\ref{fig:Fig1-Laser-scheme-noise}.
The laser cavity mirrors are equipped respectively with a slow piezo-electric (PZT) module with control bandwith $\sim$100~Hz to extend the laser's frequency tuning range up to 2.5~GHz, and a fast PZT module with control bandwidth $\sim$15~kHz for fast corrections.
We use a KTP crystal which combines two functions: intracavity electro-optic modulator (EOM)~\cite{Rusov_KTP_EOM} and birefringent filter.
The crystal axis is rotated at a small angle relative to the polarization of the radiation.
The electro-optic properties of the crystal provide a high control bandwidth of 400~kHz.
Its temperature can be used to tune the laser frequency by 120~GHz around 946~nm, while keeping an output power of $\sim$200~mW.
To have a very long operation time within the frequency tuning range of the EOM and PZT modules, the KTP crystal must be kept at a very stable temperature, which is difficult to achieve because of temperature gradients inside the laser cavity.
We use temperature sensors placed at different positions on the laser head~(figure~\ref{fig:Fig1-Laser-scheme-noise}) and find a correlation between the laser frequency and the temperature of the KTP crystal (sensor~\#0), the brass cavity enclosure (sensor~\#1) and the laser's heat sink (sensor~\#2).
We use a digital controller to measure continuously the temperature of the three sensors and apply a feed-forward algorithm with weighted coefficients determined experimentally to correct the long-term fluctuations of the tuning range.
After corrections the laser wavelength stays within 8~pm (2.5~GHz) over weeks.
The free-running frequency noise of the laser is depicted in figure~\ref{fig:Fig1-Laser-scheme-noise} and is about a factor 10 lower than the noise of laser diodes around this frequency. The measurement is performed on the side of a fringe of a Fabry-Perot cavity with a full width at half maximum of 10~MHz, which acts as a frequency-to-amplitude converter. To distinguish between frequency and amplitude noise, the laser intensity is stabilized with a double-pass acousto-optic modulator (AOM) with $\sim$700~kHz bandwidth.

\label{sec:2-Lock-cavity}

\begin{figure}[htbp]
	\centering
		\includegraphics[width=\linewidth]{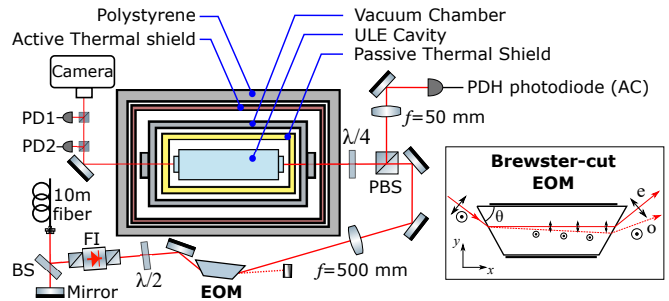}
	\caption{Optical set-up for the PDH frequency stabilization to the cavity. BS: beam splitter; FI: Faraday isolator; PBS: polarizing beam splitter; PD1 and PD2: photodiodes in transmission. We use a trapezoidal EOM crystal cut at a Brewster angle $\theta$ to obtain a large separation of the ordinary ("o") and extra-ordinary ("e") polarization components of the light, thus achieving low RAM below $2\times10^{-6}$.}
	\label{fig:Fig2-Set-up}
\end{figure}

The laser is frequency stabilized to a high finesse Fabry-Perot cavity composed of a 30~cm ultra-low expansion (ULE\textregistered) cuboid spacer~\cite{keller_simple_2014} and fused silica mirrors with high reflection coatings at 946~nm. 
The calculated relative thermal noise limit of this cavity is $1\times10^{-16}$~\cite{Kessler_2012}.
The cavity rests on a simple vibration insensitive support, which has previously been characterized with mirrors at 822~nm~\cite{keller_simple_2014}. The measured acceleration sensitivities were in the low $10^{-11}\mathrm{/ms^{-2}}$ for all directions. We exchanged the mirrors and placed them with tolerances better than 0.1~mm, so that deviations of the vibration sensitivities are expected to be below $4\times10^{-12}\mathrm{/ms^{-2}}$. 
The system is mounted on a passive vibration isolation platform, which damps vibrations above 0.7~Hz.
The cavity is placed inside a $\sim$10~mm thick aluminium vacuum chamber, surrounded by an active thermal shield made of aluminium (see figure~\ref{fig:Fig2-Set-up}), which is temperature stabilized. Passive thermal insulation is provided by a gold-coated copper shield in vacuum, a polystyrene layer around the active thermal shield and a simple wood box enclosure.
We measure the frequency response of the cavity to a 0.5°C temperature step on the active shield and derive a time constant around 91~h.
The zero-crossing of the cavity coefficient of thermal expansion is measured to be 40.4°C, very close to the simulated value of 39.6°C~\cite{Legero_2010}.
The laser frequency is stabilized to a cavity resonance via the PDH technique. The optical set-up is depicted in figure~\ref{fig:Fig2-Set-up}. The laser is placed on a different breadboard separated by a 10~m polarization maintaining (PM) fiber with active stabilization of the optical path.
After a 60~dB Faraday isolator the laser is phase modulated by an EOM at 11.9~MHz.
The residual amplitude modulation~(RAM) of EOMs is one of the important instability contributions of PDH-stabilized lasers.
It can be produced by interferences between the EOM facets~\cite{Whittaker_1985} or from polarization rotation which converts to amplitude modulation after passing through polarizing  optical elements~\cite{Zhang_2014}.
EOMs based on crystals with Brewster-cut trapezoidal~\cite{Dooley_RAM_2012} or parallelogram~\cite{Tai2016-RAM} prism shape avoid interferences within the crystal and exploit its birefringence to separate ordinary and extra-ordinary components of the light in two beams~(see figure~\ref{fig:Fig2-Set-up}), thus exhibiting very small intrinsic RAM.
We use here a home-built EOM based on a Brewster-cut trapezoidal prism KTP crystal for achieving a large separation of the two polarization components of the light.
Using a photodiode placed directly behind the EOM, we rescale the amplitude fluctuations to voltage fluctuations of the PDH error signal and estimate a short-term passive fractional RAM below $2\times10^{-6}$. 
This is more than a factor 3 improvement over the ultra-low RAM obtained with parallelogram prism shape EOM~\cite{Tai2016-RAM} and is very close to the best results achieved via more complex active stabilization~\cite{Zhang_2014}.
For our lock parameters, this contributes to a frequency instability below $2\times10^{-17}$ at 4~s.
The light reflected from the cavity is sent to a 50~MHz bandwidth photodiode. 
The transmitted light on resonance is sent to two 2.5~MHz bandwidth photodiodes for the intensity stabilization and its out-of-loop characterization.

\begin{figure}[htbp]
	\centering
	\includegraphics[width=\linewidth]{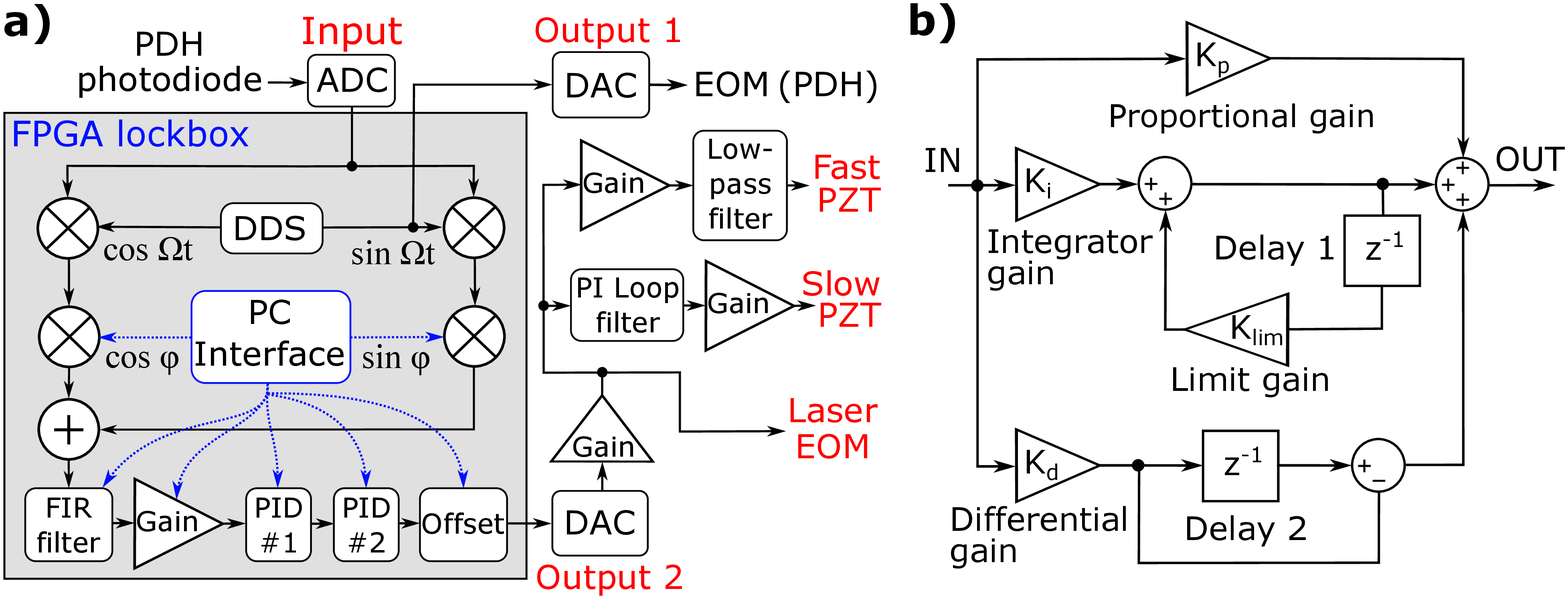}
	\caption{
		\textbf{a} - Block diagram of the locking electronics. The FPGA lockbox (grey area) contains two PID controllers with adjustable gain limit at low frequencies.
		The output signal of the lockbox controls the intracavity EOM, fast and slow PZT mirrors.
		\textbf{b} - Block diagram of one digital PID with additional limitation
		of the integrator gain.
	}
	\label{fig:Fig3-Digital-Lockbox}
\end{figure}

The laser is frequency stabilized to the cavity by a digital controller using only the signal from the PDH photodiode.
The frequency mixer, phase shifter, rf filters and PID are implemented on an EP3C25 field-programmable gate array (FPGA) from Altera including 14 bit 65~Msps analog-to-digital converters and 125~Msps digital-to-analog converters.
We program a quadrature lock-in amplifier (see figure~\ref{fig:Fig3-Digital-Lockbox}a) at 11.9~MHz with two sines and cosines produced by a direct-digital synthesizer (DDS). The DDS signal is also used to drive the EOM.
The demodulated PDH signal~\cite{Drever_1983} after the mixer passes through a finite-impulse response filter to suppress harmonics of the modulation.
Two digital PIDs are used in series for the stabilization of the laser to the cavity.
We apply an original algorithm for the PID to have an adjustable limit $K_{lim}$ of the integral gain $K_{i}$, which simplifies the optimization of the PID controllers.
The operation principle of one PID is shown in figure~\ref{fig:Fig3-Digital-Lockbox}b.
The maximum integrational gain is:
$K_i^{max} =K_{i}/(1-K_{lim})$.
The output of the digital lockbox is applied to the laser intracavity EOM, which provides the wide bandwidth of the laser frequency control.
The same amplified signal is filtered and applied to the fast PZT mirror. 
The filter suppresses the mechanical resonances of the PZT at 70~kHz.
An additional PI loop-filter is used to control the slow PZT mirror at low frequencies ($\lesssim$100~Hz).
The total loop bandwidth is $\sim$400~kHz, limited by the intracavity EOM.

\label{sec:3C-instab-contrib}

\begin{figure}[htbp]
	\centering
		\begin{minipage}{\linewidth}
			\includegraphics[width=\linewidth]{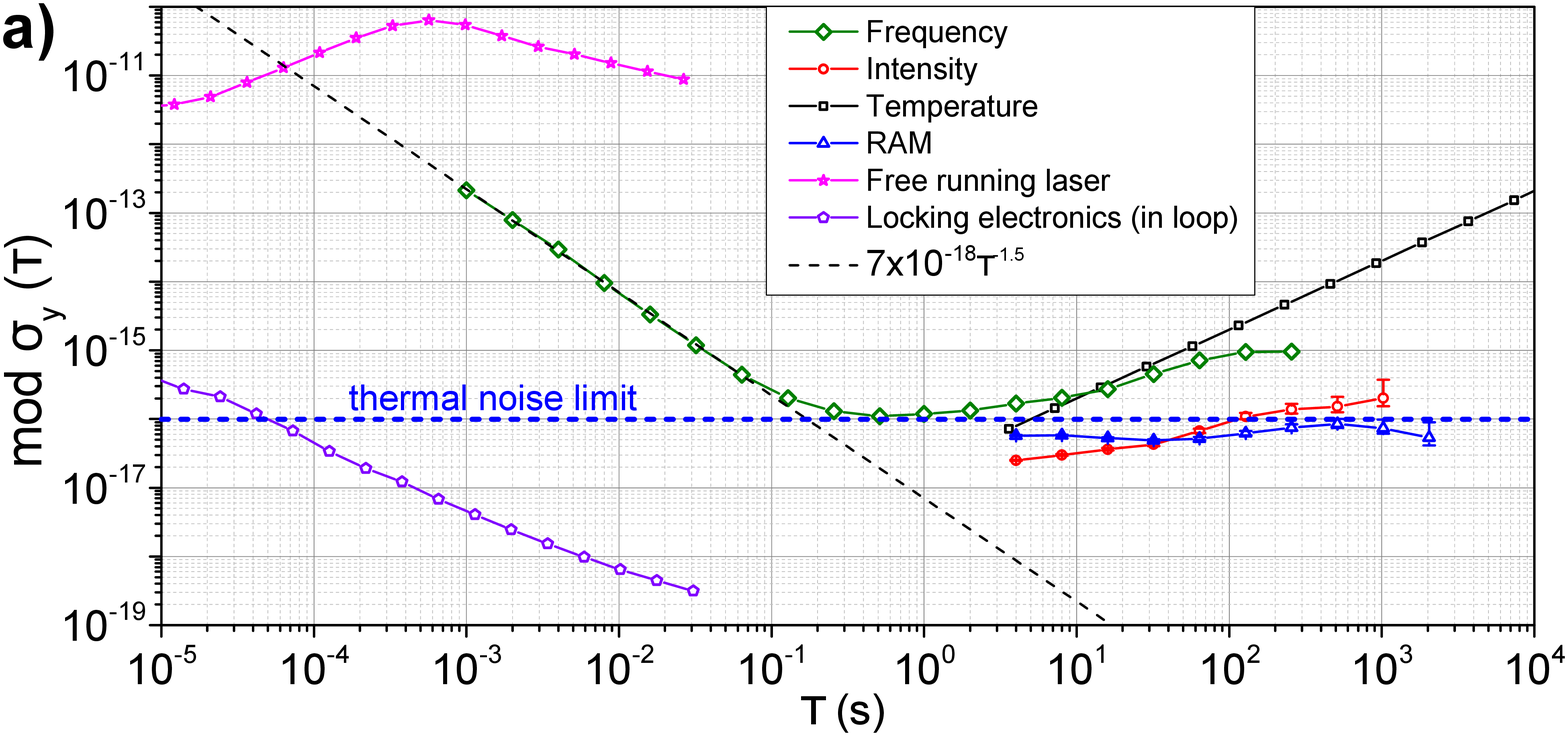}
			\includegraphics[width=\linewidth]{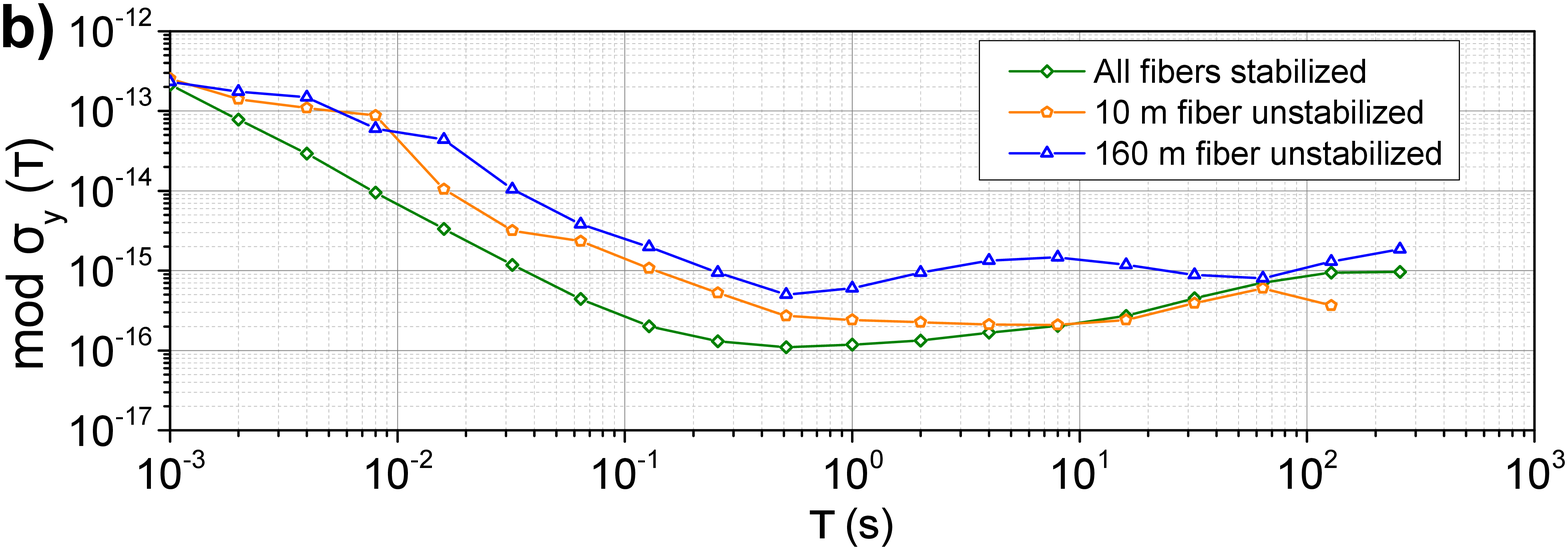}
		\end{minipage}	
	\caption{\textbf{a} - Fractional frequency instability (modified Allan deviation) of the 946~nm laser locked to the ULE cavity and the important contributions to the instability. \textbf{b} - Measured instability depending on fiber length stabilizations. The linear drift is removed for all frequency measurements in \textbf{a} and \textbf{b}.}
	\label{fig:Fig4-Instab}
\end{figure}

\begin{figure}[htbp]
	\centering
			\includegraphics[width=\linewidth]{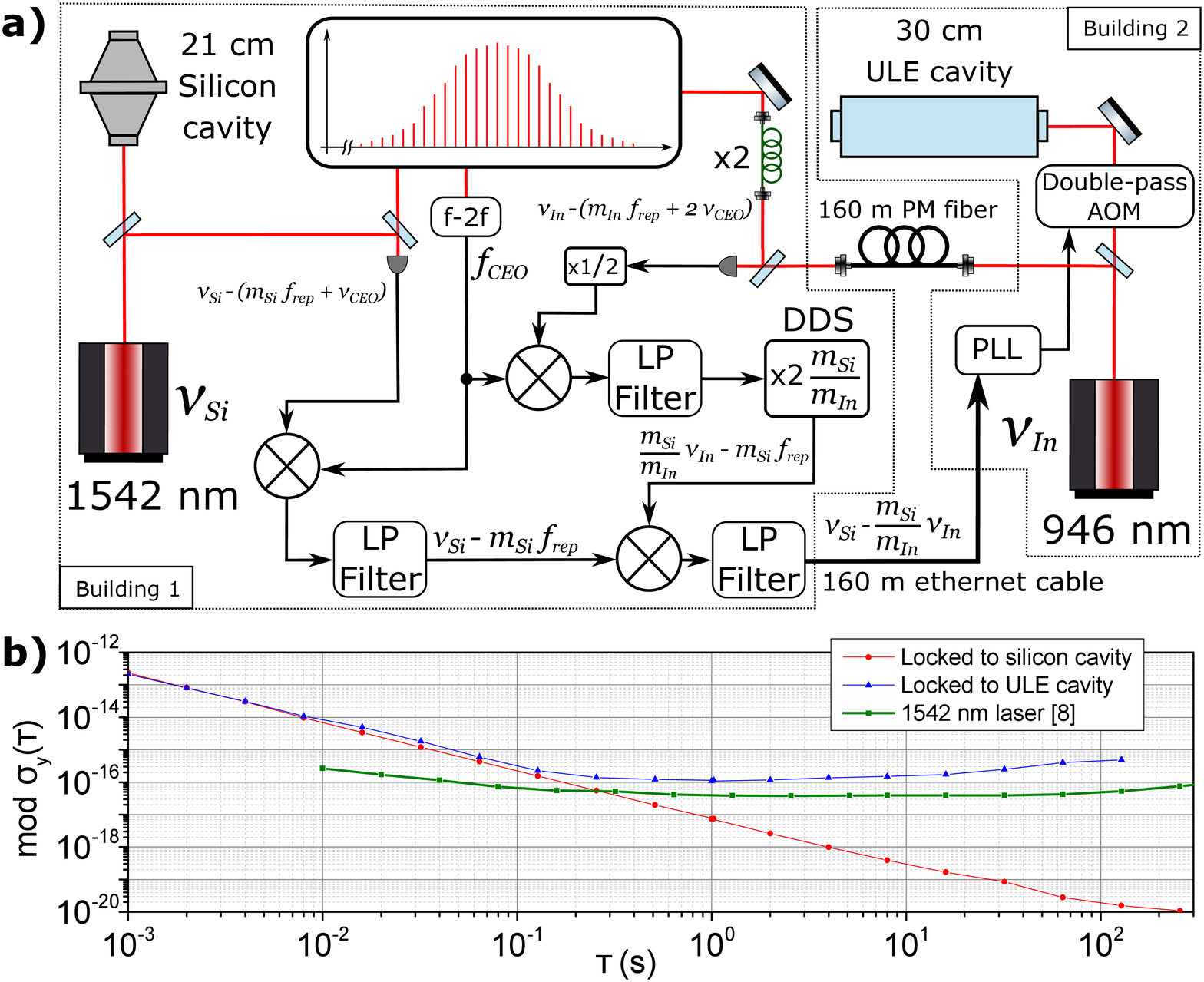}
	\caption{\textbf{a} - Scheme of the transfer lock to the silicon cavity via an optical frequency comb. A Phase-Lock-Loop (PLL) uses the transfer beat to correct the 946~nm frequency by means of a double-pass AOM. \textbf{b}~-~Fractional frequency instability (modified Allan deviation) of the comparison between the lasers when the 946~nm laser is stabilized to the ULE cavity (blue triangles) or to the silicon cavity (red dots), which corresponds to an in-loop measurement. The green squares correspond to the 1542~nm laser instability from~\cite{Matei2017}.}
	\label{fig:Fig5-Transfer_Lock}
\end{figure}

The radiation of the ultra-stable laser is sent to a frequency comb via a 160~m PM fiber with stabilized optical path (see figure~\ref{fig:Fig5-Transfer_Lock}a).
We compare the 946~nm laser with a 1542~nm laser stabilized to a cryogenic silicon cavity~\cite{Matei2017} via the frequency comb using a transfer oscillator technique introduced in~\cite{Stenger_2002}.
The optical frequency comb acts as an intermediate transfer oscillator, while its carrier-envelope offset (CEO) frequency $\nu_\mathrm{CEO}$ and repetition rate $f_\mathrm{rep}$ are allowed to fluctuate. 
These signals are eliminated in the final so-called transfer beat either via direct computation from the individual beat notes between the CW light fields and the comb or via RF signal processing in the electronics generating the transfer beat (see figure~\ref{fig:Fig5-Transfer_Lock}a).
This beat at 55~MHz is then given as $\nu_\mathrm{Si-In}=\nu_\mathrm{Si}-(m_\mathrm{Si}/m_\mathrm{In})\nu_\mathrm{In}$, with $m_\mathrm{Si}$ and $m_\mathrm{In}$ the comb line order numbers of the beat notes at 1542~nm and 946~nm respectively.
The elimination of $\nu_\mathrm{CEO}$ and $f_\mathrm{rep}$ and thus of their fluctuations is in practice limited by the common mode rejection achieved between the different paths in the transfer beat setup. Processes that can deteriorate the common mode rejection are (a)~relative uncompensated optical path length fluctuations including different branches in the frequency comb generator, (b)~limited signal-to-noise ratios (SNRs) in the photo-detection of the CEO beat and the beats between the CW fields and the comb, and (c)~excess noise in the transfer beat RF electronics, like reduction of SNR due to signal compression or jitter of the DDS implementing the frequency ratio $m_\mathrm{Si}/m_\mathrm{In}$.
The fractional frequency instability of the 1542~nm laser is $4\times10^{-17}$ between 0.8~s and $\sim30~s$, such that the beat instability is almost entirely due to the 946~nm laser.
Figure~\ref{fig:Fig4-Instab}a shows the fractional frequency instability of the 946~nm laser obtained from computation of a transfer beat from the beat notes with the comb. The linear drift of $87~\mathrm{mHz/s}$ is removed.
We reach a total fractional frequency instability of $1.1\times10^{-16}$ at 1~s, which is close to the cavity thermal noise limit of $1\times10^{-16}$. 
To estimate the contribution of intensity fluctuations to the frequency instability, we change the intracavity power in steps of $\pm$24~µW and measure the corresponding frequency shift. We derive a fractional frequency per intracavity power sensitivity of $4.2\times10^{-14}/$µW.
Using the out-of-loop photodiode placed behind the cavity, we measure a fractional intensity instability of $10^{-5}$ at 4~s, which converts to $2\times10^{-17}$ in the fractional frequency instability.
The total RAM including the contributions of the EOM, etalon effects and beam pointing on the PDH photodiode is $5\times10^{-17}$ at 4~s, which we derive from the PDH error signal out of resonance.
The cavity temperature instabilities are derived from the temperature measured on the active thermal shield, which is filtered using the measured thermal time constant of 91~h.
They are then converted to a contribution to the frequency instability via the measured cavity temperature sensitivity of $\lesssim1.2\times10^{-10}$/K.
Non-linear drifts, probably due to temperature inhomogeneities of the thermal shield, dominate the instability above 1~s.
Figure~\ref{fig:Fig4-Instab}b shows the measured instability depending on length stabilization of the fibered links. One can note that even the stabilization of the small 10~m fiber between the laser and the cavity is required to reach an instability of $10^{-16}$.

\label{sec:3-Transfer-Lock}

This level of instability is suitable for interrogation of $\sim$10~In$^{+}$~ions in the multi-ion~In$^{+}$ clock~\cite{Herschbach_2012,Keller2019-PRA}. 
Nevertheless, its theoretical quantum projection noise limit can be estimated at $7\times10^{-17}/\sqrt{\tau\mathrm{(s)}}$ for 100~In$^{+}$ ions, which surpasses the performances of the interrogation laser~\cite{keller_simple_2014}. 
We therefore implement the transfer oscillator technique presented previously to transfer the outstanding stability of the 1542~nm silicon cavity stabilized laser~\cite{Matei2017} to our 946~nm In$^+$ clock laser.
This technique was previously demonstrated in a stability transfer from the Si cavity laser to a Sr lattice clock laser~\cite{Hagemann_2013} with an instability in the range of 1~to~$3\times10^{-16}$ at 1~s.
This is also advantageous for reducing long-term instabilities due to temperature fluctuations of the cavity or the creeping of the ULE glass.
Here, the transfer beat at 55~MHz is used in a control loop with bandwidth of $\sim$160~Hz to eliminate the relative fluctuations and thus transfer the stability by means of an AOM frequency shifter as actuator for the frequency of the 946~nm In$^+$ clock interrogation field (see figure~\ref{fig:Fig5-Transfer_Lock}a).
The fractional frequency instability of the transfer beat computed from the individual beat notes with the comb is depicted in figure~\ref{fig:Fig5-Transfer_Lock}b when the 946~nm laser is only locked to the ULE cavity and when the transfer scheme is active (in-loop measurement).
The in-loop instability shows that we correctly stabilize to the silicon cavity and that the additional noise due to the control loop is still compatible with an instability of $7\times10^{-18}$ at 1~s.
We are currently limited by the signal-to-noise ratio of the beats with the comb ($\sim$-80dBc/Hz), which can be observed as a $\tau^{-1.5}$ slope in figures~\ref{fig:Fig4-Instab}~and~\ref{fig:Fig5-Transfer_Lock}b, and by excess noise in the transfer electronics.
The actual 946~nm laser instability after transfer-lock may currently be limited by uncompensated path length fluctuations between the different branches of the comb, which can lead to instability contributions as high as $2\times10^{-16}/\tau(\mathrm{s})$~\cite{Hagemann_2013}.
This will in the future be replaced by a common branch of the comb, where instabilities well below $10^{-17}$ can be achieved above 1~s~\cite{Leopardi_2017}.


In summary we have presented a laser system at 946~nm based on a home-built Nd:YAG laser with a high control bandwidth, which is stabilized to a 30~cm ULE Fabry-Perot cavity using a Brewster-cut EOM with short-term passive intrinsic RAM below $2\times10^{-6}$.
To stabilize the laser to the cavity, we developed a digital controller replacing all key components of the PDH stabilization scheme. The controller offers several advantages over classical analog schemes, such as easy configuration of the PID parameters, precision control of the phase, and the mixer is free of offset drift.
We demonstrate with our system a fractional frequency instability of $1.1\times10^{-16}$ at 1~s with long-term performance dominated by non-linear drifts probably due to temperature inhomogeneities of the outer aluminium shield, which can be improved by better thermal contacts and a more homogeneous heat load.
Our approach allows for realizing robust and easy-to-operate ultra-stable lasers at room temperature at the $10^{-16}$ level.
To reach ultimate performances, we further stabilize the 946~nm laser to a 1542~nm laser locked to a cryogenic silicon cavity exhibiting a fractional frequency instability of $4\times10^{-17}$ at 1~s.
In the future, using crystalline coatings~\cite{Cole_2016} or nanostructured mirrors~\cite{Dickmann_2018} may allow us to reach instabilities in the $10^{-17}$ range with simple room temperature systems.
\vspace{9pt}


\noindent\textbf{\Large Funding.}
\hspace{1pt}
Russian Foundation for Basic Research project 16-52-12045; German Research Foundation grant \mbox{ME~3648/1-3} and CRC~1227 (DQ-mat), project B03; 
Via the JRP 15SIB03 OC18, this project has received funding from the EMPIR programme co-financed by the Participating States and from the European Union's Horizon 2020 research and innovation programme.
\vspace{9pt}

\noindent\textbf{\Large Aknowlegment.}
The authors thank T.~Legero and U.~Sterr for providing the silicon cavity signal, M.~Skvortsov for designing and developing the low-RAM Brewster-cut EOM and S.~A.~King for careful reading of this manuscript.

\bibliography{Didier-946nm}


\end{document}